\documentclass[aps,prl,reprint,nofootinbib,superscriptaddress]{revtex4-1}
\usepackage{amssymb}

\usepackage{tabularx}
\usepackage{subfigure}
\usepackage{amsmath,amssymb}
\usepackage{latexsym}
\usepackage{psfrag}
\usepackage{epsfig}
\usepackage{graphicx,subfigure}
\usepackage{array}
\usepackage{tabularx}
\usepackage{hyperref}

\usepackage{multirow}
\usepackage{epstopdf}
\usepackage{graphicx}
\usepackage{color}
\usepackage{algorithm}
\newtheorem{deft}{Definition}

\begin{document}
\title{Fast Reconstruction of High-qubit Quantum States via Low Rate Measurements}
\author{K. Li}
\affiliation{Department of Automation, University of Science and Technology of China, Hefei, 230027, China}%
\affiliation{Imperial College London, MRC Institute of Medical Sciences, London, W12 0NN, UK}
\author{J. Zhang}
\affiliation{Department of Automation, University of Science and Technology of China, Hefei, 230027, China}
\author{S. Cong}
\email{scong@ustc.edu.cn}
\affiliation{Department of Automation, University of Science and Technology of China, Hefei, 230027, China}

\begin{abstract}
Due to the exponential complexity of the resources required by quantum state tomography (QST), people are interested in approaches towards identifying quantum states which require less effort and time. In this paper, we provide a tailored and efficient method for reconstructing mixed quantum states up to $12$ (or even more) qubits from an incomplete set of observables subject to noises. Our method is applicable to any pure or nearly pure state $\rho$, and can be extended to many states of interest in quantum information processing, such as multi-particle entangled $W$ state, GHZ state and cluster states that are matrix product operators of low dimensions. The method applies the quantum density matrix constraints to a quantum compressive sensing optimization problem, and exploits a modified Quantum Alternating Direction Multiplier Method (Quantum-ADMM) to accelerate the convergence. Our algorithm takes $8,35$ and $226$ seconds respectively to reconstruct superposition state density matrices of $10,11,12$ qubits with acceptable fidelity, using less than $1 \%$ of measurements of expectation. To our knowledge it is the fastest realization that people can achieve using a normal desktop. We further discuss applications of this method using experimental data of mixed states obtained in an ion trap experiment of up to $8$ qubits.
\end{abstract}

\maketitle

\section{I. INTRODUCTION}

As quantum technologies grow rapidly in laboratories, the demand for a reliable and practical quantum state tomography of prepared states is high for estimating systems of larger numbers of qubits \citep{Ariano-QuantumTo, Lvovsky-Continuous,Cramer-EffQuanStaTomo}. QST becomes a significantly important standard for verification in many quantum tasks \cite{Giova-QuantumEnhanMea-Science, Schwemmer-ExpComEff}. It is known that when the set of experiments is informationally (over) complete, the state of physical systems can be uniquely determined and described as a density matrix $\rho$ \cite{Bergou-QuanStaEst}. The conventional tomography requires resource-intensive scaling to large system due to the inherent dimensionality problem, namely the exponential growth of the $n$-qubit in the Hilbert space \cite{Shabani-EffiMeaOfQuan,Lloyd-QuantumPri}, which is deemed a barrier to extending QST to higher-qubit scenarios. Many distinguished works have been done in this field, which achieved the reconstruction of a large number of qubits \cite{Cramer-EffQuanStaTomo,Riofrio-ExpQuanCSforSeven,Negreve-BenQuanContMet}, or reduced the complexity of algorithm \cite{flammia2011direct,baumgratz2013scalable,smolin2012efficient}, under various (or without) prior information. Using fewer measurements and simpler methods to reconstruct large scale quantum states remains a challenge for physicists and engineering scientists.

As a novel signal processing technique, compressive sensing (CS) has been implemented in QST in both theory \cite{donoho-cs,Gross-QuantumState,Flammia-QuantumTomo} and practice \cite{Cramer-EffQuanStaTomo,Riofrio-ExpQuanCSforSeven}. CS exploits the structure information of density matrices (e.g. high purity) in reconstruction so that merely incomplete information is needed for accurately recovering $\rho$ \cite{Liu-Exp-Quantum,Li-ARobust}.
In this paper we use the CS technique to reduce the sampling rate and develop a new algorithm to reconstruct quantum states more efficiently.
Specifically, a simple iterative algorithm, called Quantum-ADMM is proposed by applying quantum constraints (e.g. Hermitian, trace) to the ADMM framework, an increasingly popular method in optimizations. The algorithm projects the objective density matrix to the measurement function and quantum constraints alternately, and significant modifications have been made accordingly to make it fit for complex quantum computations. The proposed algorithm has been verified on simulated data, showing that it is capable of reconstructing a $12$-qubit system in pure states (or nearly pure mixed states) with the fastest computation to date, and it can be easily extended to larger systems. Simulations using experimental data obtained in an ion trap experiment is carried out, followed by a discussion compared to other state-of-the-art approaches.

\section{II. COMPRESSIVE QUANTUM TOMOGRAPHY}
Consider a system consisting of $n$ qubits, and its density matrix $\rho$ is uniquely described as a $d \times d$ matrix where $d=2^n$. Normally, the observables in quantum mechanics are Hermitian operators, and the expectation value of the Hermitian operator $\mathbf{\omega}_i, i = \{1, \cdots, d^2\}$ applied to a quantum state $\rho$ is measured as
\begin{equation}\label{eq:y_i}
y_i = Tr(\mathbf{\rho\omega}_i).
\end{equation}

As most quantum compressive sensing papers assume, we use the expectation $y_i$ as measurements of the system \cite{Cramer-EffQuanStaTomo,Riofrio-ExpQuanCSforSeven,flammia2011direct,Flammia-QuantumTomo}. The Hermitian operators $\mathbf{\omega}_i$ are a series of orthogonal bases, such as (but not restricted to) tensor products of Pauli matrices $\{\sigma_0,\sigma_1, \sigma_2, \sigma_3\} = \left\{ \left( \begin{array}{cc}1 & 0 \\ 0 & 1 \end{array} \right), \left(\begin{array}{cc}0& 1 \\ 1 & 0 \end{array}  \right), \left(\begin{array}{cc}0 & -i \\ i & 0 \end{array} \right) ,\left(\begin{array}{cc}1 & 0 \\ 0 & -1 \end{array}   \right)\right\}$. We assume and apply a rank-$r$ constraint on the density matrix, $r \ll d$. It has been rigorously shown that $M = O(rd \log^2 d) \ll d^2$ experimental measured parameters are sufficient to recover a rank-$r$ $\rho$ even when the eigenbasis is unknown as long as that rank RIP is satisfied with overwhelming probability \cite{Gross-QuantumState}. Our low-rank estimation can be appropriate in general cases, because statistical noise often allows large eigenvectors to be reliably reconstructed, while remaining unimportant eigenvectors behave in a way consistent with random matrices \cite{Riofrio-ExpQuanCSforSeven}. Hence in our model, we rewrite (\ref{eq:y_i}) in a matrix form after random sampling $M$ out of $d^2$ measurements subject to Gaussian noises:
\begin{equation}\label{eq:y=Arho+e}
\mathbf{y} = \mathbf{A} \text{vec}(\rho) + \mathbf{e},
\end{equation}
where $\mathbf{y} \in \mathbf{C}^{M \times 1}$ is the measurement vector of expectations, $\mathbf{A} \in \mathbf{C}^{M \times d^2}$ represents the matrix form of sampling operator $\mathcal{A}(\rho)=(Tr(\rho \omega_1), \cdots , Tr(\rho \omega_M))^T: \mathbf{C}^{ d \times d} \rightarrow \mathbf{C}^{M }$, $\text{vec}(\cdot)$ is the vectorize operator, and $\mathbf{e}$ denotes the $0$-mean noise subject to $\rho$. Given the rank-$r$ and quantum constraints on $\rho$, we pursue the solution of the following optimization problem:
\begin{equation}\label{eq:rho+I}
 \min_{\rho} ||\rho||_* + I_C(\rho), \ \ \text{s.t.} \  ||\mathbf{A} \text{vec}(\rho)- \mathbf{y}||_2^2 \leq \delta,
\end{equation}
where $||\cdot||_*$ denotes the nuclear norm, $||\rho||_* = \sum s_i$, $s_i$ is the singular value of $\rho$; $\delta >0$, $I_C(\rho)$ is the indictor function as the quantum constraints on a convex set $C$. Here without loss of generality, we set $I_C(\rho) = \left\{ \begin{array}{cc} 0,  & \text{if} \ \ \rho^* = \rho, \rho \succeq 0 \\ \infty, & \ \text{otherwise} \end{array} \right.$. $\rho^*$ denotes the conjugate transpose of $\rho$. The function of $I_C(\rho)$ is projecting $\rho$ to a Hermitian matrix.

\section{III. APPLYING Q-ADMM TO RECONSTRUCTION}
ADMM is an old technique in optimization proposed by Gabay etc. in 1970s \cite{gabay1976dual}. It was redeveloped by Boyd et al. in control engineering \cite{Boyd-DistributedOpt}. It divides complex optimization problem to separate steps, pursues the best solution alternately and finally finds the convergence. One can refer to the supplementary materials for the framework of ADMM. In our problem, we formulate (\ref{eq:rho+I}) into two objectives: low-rank and reducing errors, by introducing an auxiliary variable $\mathbf{e} \in \mathbf{C}^M$:
\begin{equation}\label{eq:rho+I+e}
 \min_{\rho} \gamma ||\rho||_* + I_C(\rho) + 1/2 ||\mathbf{e}||_2^2,  \ \text{s.t.} \mathbf{A} \text{vec}(\rho) + \mathbf{e} = \mathbf{y}.
\end{equation}
Here, we choose the augmented Lagrangian of (\ref{eq:rho+I+e}) as (\ref{eq:rho+I+e+Lagr}) (see the top next page).
\begin{figure*}[th]
\centering
\begin{equation}\label{eq:rho+I+e+Lagr}
 \min_{\rho} \gamma ||\rho||_* + I_C(\rho) + 1/2 ||\mathbf{e}||_2^2 + \langle \mathbf{b}, \mathbf{A}\cdot \text{vec}(\rho) +\mathbf{e}- \mathbf{y} \rangle + \lambda/2||\mathbf{A} \cdot \text{vec}(\rho) + \mathbf{e} -\mathbf{y}||_2^2.
\end{equation}
 \hrulefill
\end{figure*}
In (\ref{eq:rho+I+e+Lagr}), $\mathbf{b} \in \mathbf{R}^M$ is the Lagrangian multiplier, $\lambda>0$ is the penalty parameter. Then an Iterative Shrinkage-Thresholding Algorithm  (ISTA) is employed to the equation. Specifically, the derivation can be separated into three steps:

step1: fix $\rho = \rho^k$ and $\mathbf{b}= \mathbf{b}^k$, (\ref{eq:rho+I+e+Lagr}) is a quadratic function with respect to the auxiliary variable $\mathbf{e}$. We impose the differential equaling zero, then
\begin{equation}
\mathbf{e}^{k+1} = (\gamma \lambda/(1 + \gamma \lambda))(-\mathbf{b}^k/\lambda-(\mathbf{A} \text{vec}(\rho^k)-\mathbf{y})),
\end{equation}
where $\rho^k$ represents the $\rho$ in the $k$th iteration.

step2: fix $\mathbf{e} = \mathbf{e}^{k+1}$, minimization of (\ref{eq:rho+I+e+Lagr}) with respect to $\rho$ is equivalent to
\begin{equation}
\min_{\rho} \gamma||\rho||_* + I_C(\rho) +\lambda/2||\mathbf{A} \cdot \text{vec}(\rho) + \mathbf{e}^{k+1} -\mathbf{y}+\mathbf{b}^k/\lambda||_2^2.
\end{equation}
We introduce ISTA here to derive an intermediate matrix $\mathbf{C}_1^{k+1}$. Since the nuclear norm is non-smooth but $l_2$ norm is, and it has a Lipschitz continuous gradient \cite{iter-soft,beck2009fast}.
\begin{equation}
\mathbf{C}_1^{k+1} = \rho^k - t^k \text{mat}(\mathbf{A}^*(\mathbf{A} \cdot \text{vec}(\rho^k)+\mathbf{e}^{k+1} - \mathbf{y} + \mathbf{b}^k/\lambda)),
\end{equation}
where $t^k>0$ is an adaptive step size of the gradient descent in the $k$th iteration. Afterwards we project $\mathbf{C}_1^{k+1}$ to the Hermitian space $\mathbf{C}_1^{k+1} = 1/2(\mathbf{C}_1^{k+1} +(\mathbf{C}_1^{k+1})^* )$. In addition, a singular value contraction operator $D_{\tau}$ is employed on $\mathbf{C}_1^{k+1} $
\begin{equation}\label{eq:SVD_tau}
\rho^{k+1} = D_{\tau}(\mathbf{C}_1^{k+1} ),
\end{equation}
where $D_{\tau}(\mathbf{X}) = \mathbf{U}\mathbf{S}_{\tau} \mathbf{V}^T $, $\mathbf{US}\mathbf{V}^T$ is the singular value decomposition of $\mathbf{X}$, $[S_{\tau}]_{i,j} = \left\{ \begin{array}{cc} x_{ij}-\tau, & \text{if}\  x_{ij} > \tau \\ x_{ij}+\tau, & \text{if} \ x_{ij} < - \tau \\ 0, & \text{otherwise}\end{array}\right.$ is a piecewise operator on individual matrix element. The positive definite and trace constraints are also employed in this step.

step3: fix $\mathbf{e} = \mathbf{e}^{k+1}$ and $\rho = \rho^{k+1}$, we update the multiplier $\mathbf{b}$
\begin{equation}\label{eq:b}
\mathbf{b}^{k+1} = \mathbf{b}^k + \kappa \lambda(\mathbf{A} \text{vec}(\rho^{k+1})+\mathbf{e}^{k+1}- \mathbf{y}),
\end{equation}
where $\kappa>0$ is a parameter relates to the convergence rate.   $\blacksquare$

In summary, the tailored ADMM iterates as follows
\begin{equation}\label{eq:ADMM_all}
\left\{ \begin{array}{l}
\mathbf{e}^{k+1} = (\gamma \lambda/(1 + \gamma \lambda))(-\mathbf{b}^k/\lambda-(\mathbf{A} \text{vec}(\rho^k)-\mathbf{y})),\\
\rho^{k+1} = D_{\tau}(\mathbf{C}_1^{k+1} ),\\
\mathbf{b}^{k+1} = \mathbf{b}^k + \kappa \lambda(\mathbf{A} \text{vec}(\rho^{k+1})+\mathbf{e}^{k+1}- \mathbf{y}).
 \end{array}\right.
\end{equation}
There are $4$ adjustable parameters in (\ref{eq:ADMM_all}): step size $t$ for gradient descent method; update step $\kappa$ for Lagrange multiplier; weight $\gamma$ that balances the low-rank and error terms; penalty parameter $\lambda$. They will be discussed later in the discussion section.

\section{IV. EXPERIMENTS}
In this part tensor products of Pauli matrices are utilized to construct the square measurement matrix and $\mathbf{A}$ in (\ref{eq:y=Arho+e}) is a sub-matrix of it generated by randomly selecting rows. Let the reconstructed state be $\hat{\rho}$ and true state be ${\rho}$, normally there are 2 criteria to measure the reconstruction performance. They are \emph{Hilbert Schmidt norm different} [6],
\begin{equation}
D({\rho},\hat{\rho}) = \frac{||\hat{\rho} -\rho ||_2^2}{||\rho||_2^2},
\end{equation}
and \emph{fidelity} \cite{flammia2011direct},
\begin{equation}\label{eq:fidelity}
\text{F}({\rho},\hat{\rho}) = Tr\left[ \sqrt{(\sqrt{\rho} \hat{\rho} \sqrt{\rho})}\right].
\end{equation}
Here we adopt both to measure the reconstruction performance. In fact $D({\rho},\hat{\rho})$, $\text{F}({\rho},\hat{\rho})$ values are very close.

\begin{figure}[t]
\centering
\includegraphics[width=7.1cm]{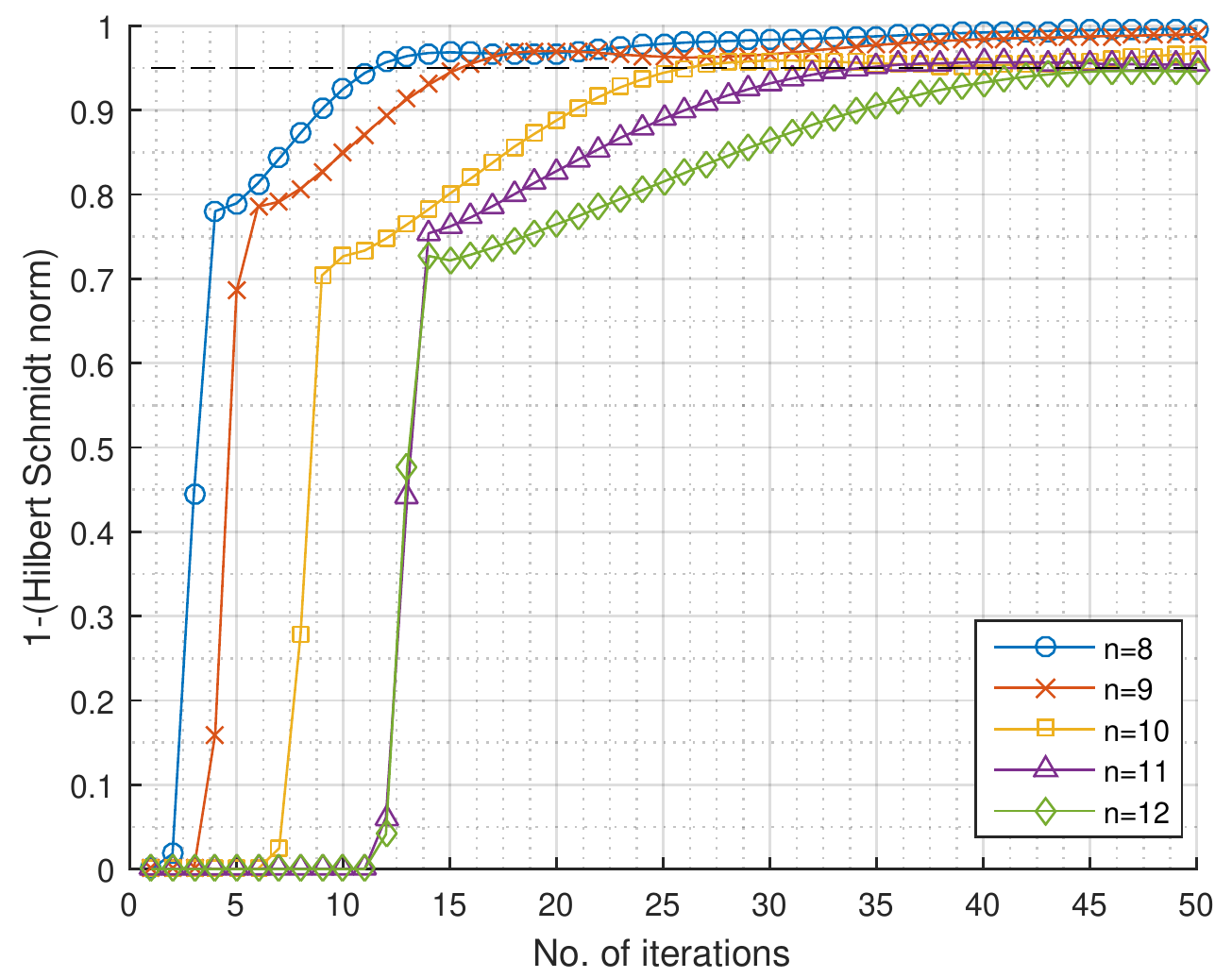}
\caption{(color online) The reconstruction performances of $n=8 \sim 12$ qubits using the proposed method are shown in terms of increasing number of iterations in different colors. The x-axis represents the number of iterations and the y-axis represents $1- D({\rho},\hat{\rho})$.  The dash line represents the $94.5 \%$ accuracy in terms of the Hilbert Schmidt norm different. Each number in the figure is the average of $100$ simulations. All curves reach an accuracy of $1- D({\rho},\hat{\rho})$ above $94.5 \%$ within $50$ iterations. Corresponding fidelity values can be referred to Table \ref{tab:1}. }
\label{fig:1}
\end{figure}

In this part, we implement our method to quantum systems with $8$-$12$ qubits, and then compare the consuming time to previous results. We use the Dell desktop with Inter Core i7-4790 CPU @3.60GHz with 16 GB RAM. The scripts are written and run using MATLAB. The true $\rho$ is generated from normalized Wishart random matrices with form as \cite{Zyczkowski-GeneratingRan} $\rho = \frac{\Psi_r \Psi_r^*}{Tr(\Psi_r \Psi_r^*)},$
where $\Psi_r$ is a complex $d\times r$ matrix with i.i.d. complex random Gaussian entries. The denominator is constructed due to the trace $1$ constraint of the density matrix. Without loss of generality, $r$ is set to $1$ making $\rho$ to be an arbitrary pure/superposition state ($r>1$ can be derived in a similar approach). Parameter values adopted in experiments are: $t=0.9$, $\kappa=1.099$, $\gamma = 1e(-4)$; $\lambda = 8,14,30,30,30$ when $n = 8,9,10,11,12$, respectively. With sampling operator generated from Pauli matrices, the measurement rate $\eta = M/d^2 \sim O((r \log^2 d)/d)$. When $r=1$, $\eta \geq \log( d)/((1+\vartheta)d)$ can recover the unique and accurate $\hat{\rho}= \rho$ with probability $P_s \geq d^{-\vartheta^2/2\ln 2(1+\vartheta/3)}$. After calculation, here we let $\vartheta=0.05$ and use $\eta = 2.98\%, 1.67\%,0.93\%, 0.51\%$ respectively to achieve a reconstruction probability larger than $98\%$. Matrices $\mathbf{A}$ for $n=8 \sim 12$ are generated as a sparse matrix in advance. The noises are added with an amplitude $\text{SNR} =40 \text{dB}$.  The reconstruction performances are demonstrated in Fig. \ref{fig:1}\footnote{Please refer to \url{https://github.com/KezhiLi/Quantum_ADMM} for codes.}. Full results are shown numerically in Table \ref{tab:1} in terms of the fidelity and reconstruction time.

\begin{table}[b]
\centering
\begin{tabular}{|c|c|c|c|c|c|}
\hline
Qubit $n$& $n=8$ & $n=9$ & $n=10$  & $n=11$  & $n=12$\\
\hline
Measurement rate $\eta$ & 3\% & 1.7\% & 1\% & 0.6\% & 0.3\%  \\
\hline
Fidelity & 0.991 & 0.988 & 0.987 & 0.986 & 0.985  \\
\hline
Number of iterations & 12 & 16 & 27 & 35 & 46 \\
\hline
Reconstruction time(s) & 0.59 & 1.78 & 7.95 & 35.03 & 226.43  \\
\hline
\end{tabular}
\caption{(color online) Table to compare the reconstruction results in terms of increasing number of qubits. The number of iterations, time and fidelity values are recorded or calculated from (\ref{eq:fidelity}) once $D({\rho},\hat{\rho})$ reaches above $94.5 \%$ accuracy. }\label{tab:1}
\end{table}

 In Table \ref{tab:1} the fidelity values are all above $0.98$ which indicate an accurate reconstruction. With the growth of qubits, the algorithm needs more number of iterations to achieve the reconstruction; however the measurement rates $\eta$ are decreasing, suffice to the compressive sensing theory [15], that the required sampling rates decrease when the number of qubits increases. The advantage of proposed algorithm is its efficiency. We only need $2,8,35,226$ seconds to recover a quantum state of $n=9,10,11,12$ qubits respectively. These are considered as the fastest to date on a single core normal desktop.

\begin{figure}[t]
\centering
\includegraphics[width=7cm]{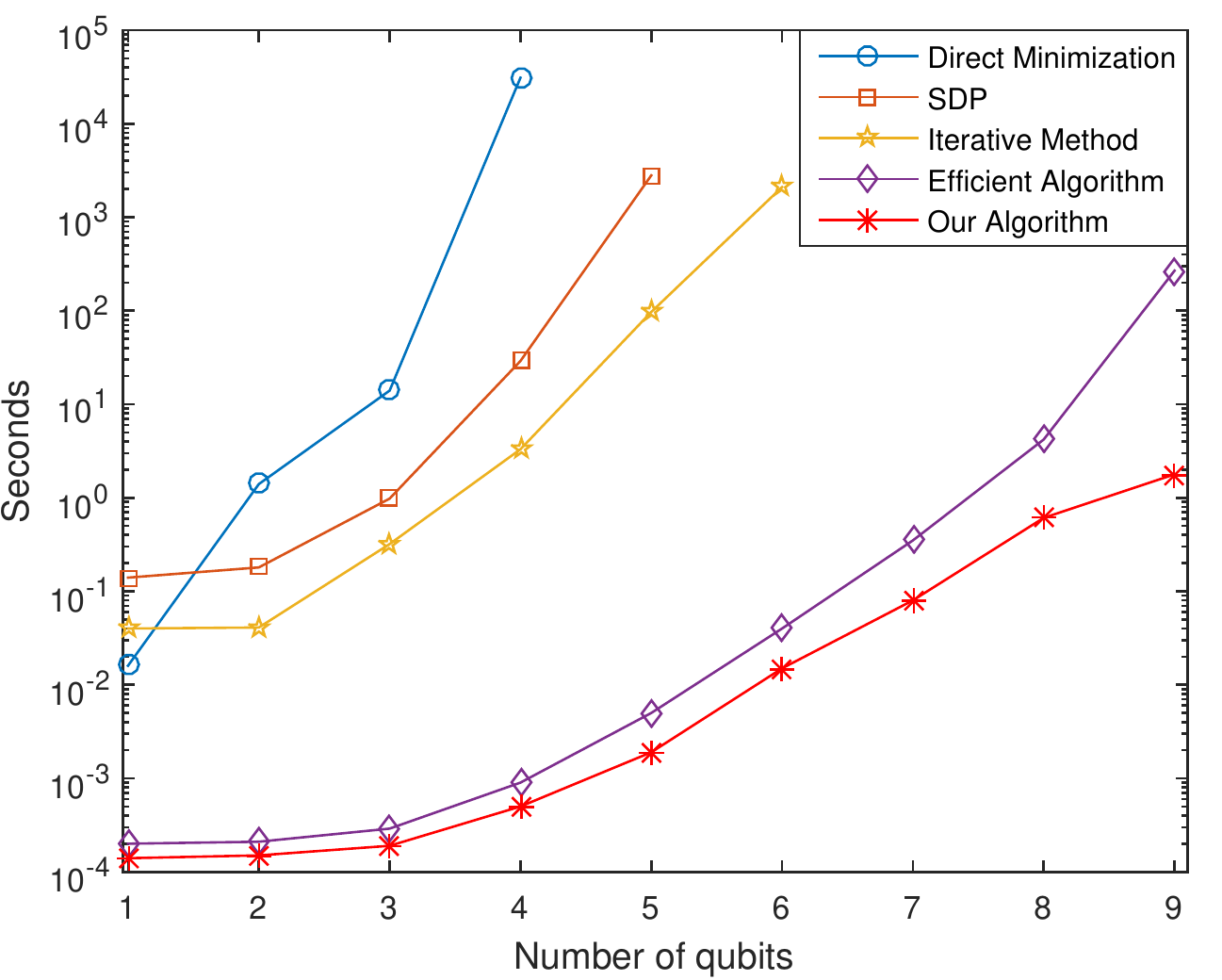}
\caption{Run time for reconstruction
of random $n$-qubit pure states subjected to Gaussian noise on Pauli measurements.
We compare four techniques: The circle points are
MATLAB’s fminsearch minimizing $Tr[(\hat{\rho}-\rho)]$ directly.
Timings for a semidefinite programming
method (SeDuMi) \cite{Sturm-UsingSeD}, the realization the iterative method of \cite{vrehavcek2008tomography} and the \emph{efficient algorithm} \cite{smolin2012efficient} are denoted as square, star and diamond points, respectively. Our algorithm is shown with $*$. All timings were performed on a single core of a 3.6 GHz Intel i7-4790 CPU in MATLAB.}\label{fig:2}
\vspace{-1.2\baselineskip}
\end{figure}

Next, we compare our algorithm to a previous method developed in \cite{smolin2012efficient} by reconstructing random $n$-qubit pure states subject to the Gaussian noise. The general settings are similar, so the two papers' results are comparable, though much less measurements are used for reconstruction in the proposed method. The \emph{efficient algorithm} developed in \cite{smolin2012efficient} is claimed as one of the fastest methods which completes a $8$-qubit reconstruction in seconds. The timings are shown in Fig. \ref{fig:2} explicitly. From Fig. \ref{fig:2} it indicates that our algorithm is the most efficient algorithm shown in the comparison, including the \emph{efficient algorithm}, particularly when the number of qubits is large.

Finally, we apply our method to experimental data. Numerical results are demonstrated at the hand of $W$ states having $8$ qubits created in an ion trap experiment \cite{haffner2005scalable}, i.e.
 \begin{equation}
 |W(\phi) \rangle = [|0...01 \rangle + e^{i\phi_1}|0... 10 \rangle + \cdots + e^{i\phi_{n-1}}|1...00\rangle]/\sqrt{n}.
 \end{equation}

\begin{figure}[t]
    \centering
    \begin{minipage}[t]{0.49\textwidth}
        \includegraphics[width=\textwidth]{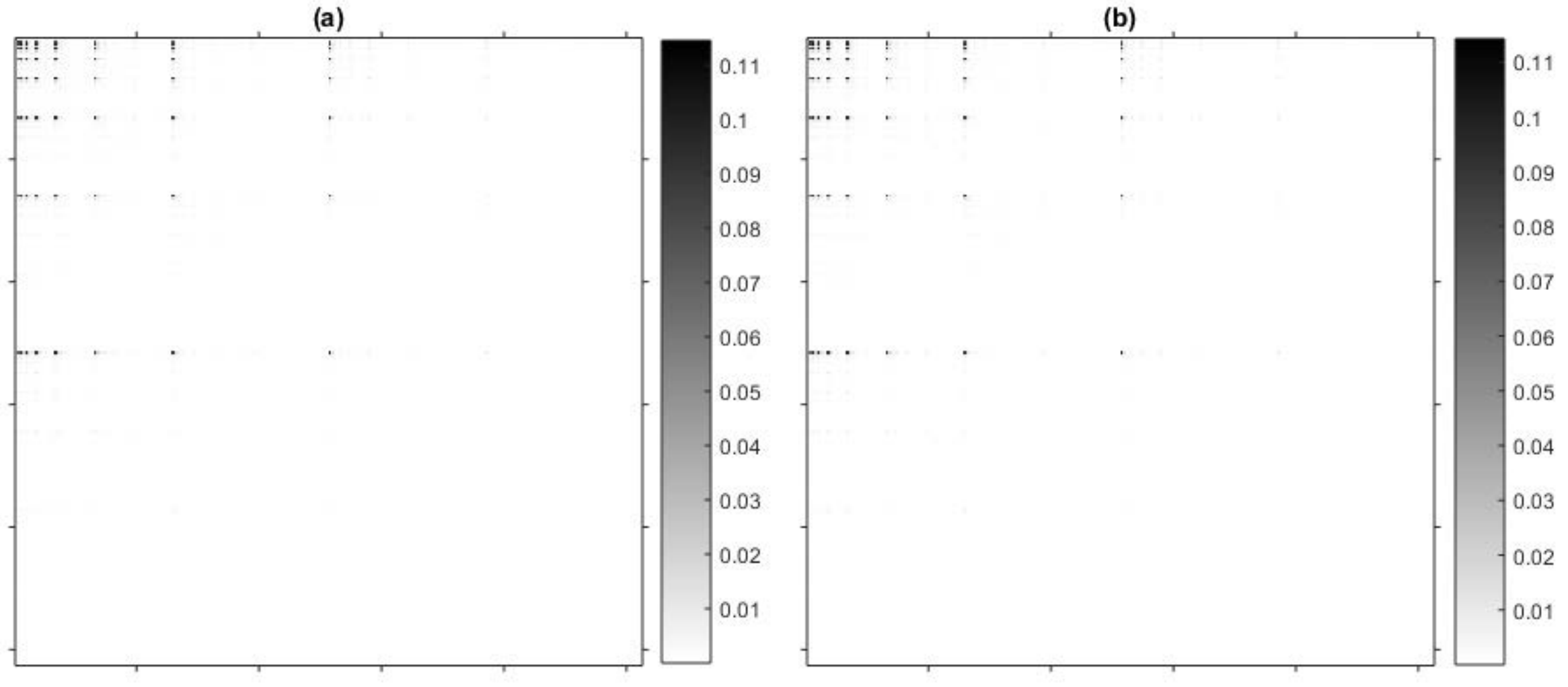}
        \caption{Absolute value of corresponding reconstructed density matrix of the experimentally realized W state. (a) Maximum likelihood estimate of full quantum state tomography $|\rho_{ML}|$ \cite{haffner2005scalable} (b) Reconstruction $|\hat{\rho}| $ using the method described in this Letter with sampling rate $\eta = 15\%$ obtained after 3 iterations, $0.14$ seconds.   }\label{fig:3-11}
    \end{minipage}
    \begin{minipage}[t]{0.49\textwidth}
        \includegraphics[width=8.2cm]{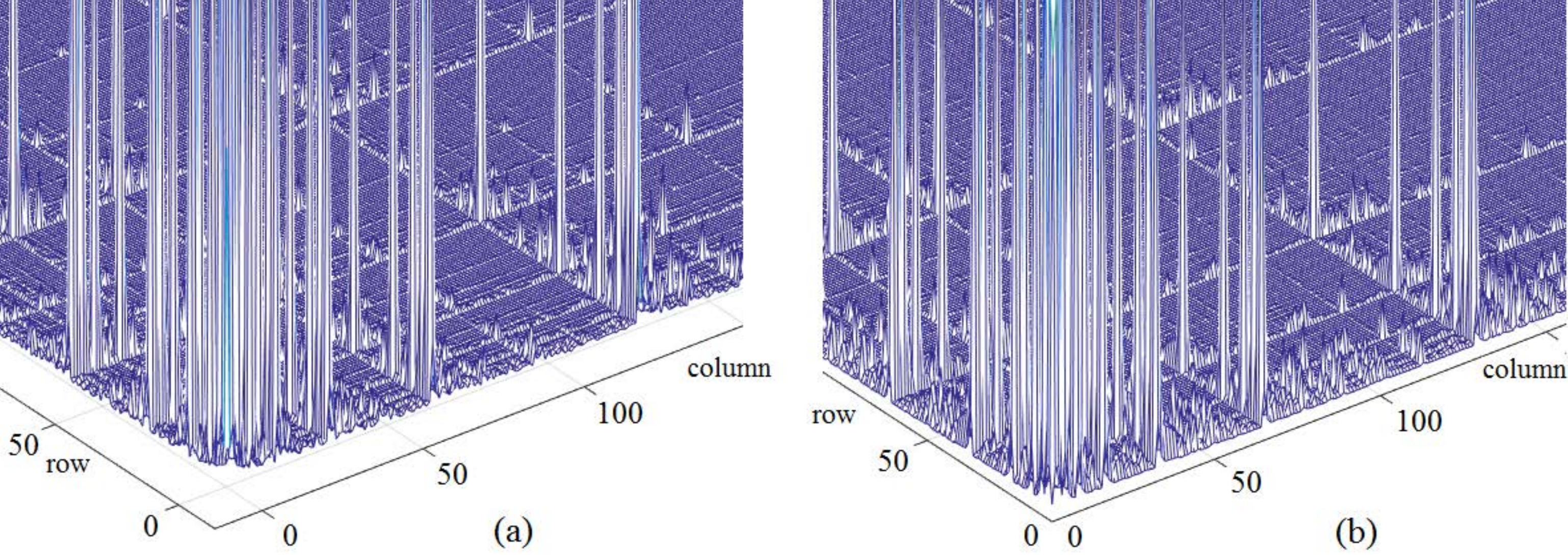}
        \caption{The comparison of magnitudes of elements in density matrices of $|\rho_{ML}|$  and  $|\hat{\rho}| $ shown in Fig. \ref{fig:3-11}. (a) The magnitudes of $|\rho_{ML}|$ (b)  The magnitudes of $|\hat{\rho}|$. The algorithm can also recover details approaching full tomography ML result.}\label{fig:3-12}
    \end{minipage}
\end{figure}

 The reconstructed result obtained in the full tomography procedure using maximum likelihood estimate (MLE) is denoted as $\rho_{ML}$. The objective state is no longer pure, which belongs to entangled states. The input to the reconstruction method is a random subset of the relative frequencies corresponding to the measurements on all subsystems (expectation value) with $\eta = 15\%$, which can be obtained in advance. A graphical representation of the reconstruction of density matrices' absolute values is in Fig. \ref{fig:3-11}, which compares our reconstructed $\hat{\rho}$ (b) to $\rho_{ML}$ (a). We achieve the renormalized Hilbert-Schmidt norm difference $D(\rho_{ML},\hat{\rho}) \leq 0.046$ after $0.14$ seconds and $0.024$ after $0.7$ seconds, with partial details shown in Fig. \ref{fig:3-12} (though there are many noises). With respect to the local phases of a pure $W$ state yields $f = \langle W(\phi_{\text{opt}})|\hat{\rho}|W(\phi_{\text{opt}}) \rangle = 0.722$ by maximizing the fidelity of the MLE \cite{haffner2005scalable,baumgratz2013scalable}. In our case we achieve a fidelity $f=0.719$ with respect the optimal $W$ state stems from the same $|W(\phi_{\text{opt}}) \rangle$ as in \cite{haffner2005scalable}. It verifies the effectiveness of algorithm under a very noisy environment, in addition to indicate that it can achieve a reconstruction approaching MLE obtained from full tomography but with lower rate samples.


\section{V. DISCUSSION}
1. This paper addresses the quantum state reconstruction problem up to $12$ qubits using a normal desktop. More qubits and faster computation can be carried out using multi-core workstations and GPU acceleration. The advantage of our method is faster reconstruction given lower rate measurements. According to the CS theory, the sampling rate can be lower when the number of qubits is larger. Further, the numerical simulations in this paper reveal this characteristic in Table \ref{tab:1} for pure states, which relieves the exponential expenses $O(d^2)$ to near linear $O(rd \log^2 d)$ \cite{Riofrio-ExpQuanCSforSeven,Liu-Universal-Pauli,Zheng-ARecAlgForCQT}. Moreover, researchers also argue that the low-rank estimates can be appropriate in the general case due to the random matrix theory \cite{Liu-Universal-Pauli}. This theory extends the application of scope of our method from pure or nearly pure states to broader states in general.

2. We assume that input of the algorithm is the expectation values of observables. This assumption is a prior condition widely present in most compressive QST works \cite{Cramer-EffQuanStaTomo, flammia2011direct,smolin2012efficient,Liu-Universal-Pauli,Li-ARobust,Liu-Exp-Quantum}. Some settings, such as Nuclear magnetic resonance (NMR), capture the expectation values directly from experiments \cite{Riofrio-ExpQuanCSforSeven,baumgratz2013scalable,Liu-Exp-Quantum}, thus this assumption is reasonable in QST.

3. With regard to the complexity of the algorithm, the slowest step is the step that solves the eigensystem in (\ref{eq:SVD_tau}), which is $O(d^3)$. Other steps are less complex thus the overall complexity is $O(d^3)$. The prior basis transformation step costs $O(d^4)$, yet it can be computed in advance before running the algorithm. The actual processing time also depends on the solver implemented, eg. we utilize the 'rsvd' function (random SVD, a fast computation of the truncated SVD) instead of 'svd' to accelerate the decomposition \cite{Halko-FindingStr}. The proposed method is designed for reconstructing pure or nearly pure states. If we know that the objective state is pure, this prior information can be adopted in the shrinkage step (\ref{eq:SVD_tau}), so that a small number of singular values can be reserved in each iteration.

4. There are several parameters in the algorithm that need to be determined. Generally speaking, parameters are determined based on experiences.
Specifically, we set $\tau = t/\lambda$, where the adjustable parameter $\lambda$ is a parameter to balance the quadratic and rank terms in the optimization. We set $\lambda = 2M/\text{norm}(\mathbf{b}), t = 1$ initially. $\lambda$ has the same function as the parameter in a standard ADMM framework \cite{Boyd-DistributedOpt}.
 $\tau \in (0,1)$ is the shrinkage parameter that determines the shrinkage step relying on the distribution of singular values of the density matrix.
 Usually for pure states $\rho$, $\tau$ can be set larger than it for non-pure states.
$\kappa \sim 1$ is a parameter to control the residual update rate and tuning $\kappa$ within the range of $(0, (\sqrt{5}+1)/2)$ often helps to improve the convergence speed.
 In addition, $\mathbf{b}^k$ is seen as the residual. We use the norm of $\mathbf{b}^k$ as the stopping criterion and compare it with a stop threshold to decide when the algorithm stops. In the experiments we set the stop threshold as $1e(-6)$, which allows $D({\rho},\hat{\rho})$ to reach above $94.5 \%$ in $100$ iterations .

5. The convergence of the ADMM algorithms in quantum state tomography is discussed and proved explicitly in our other works. Please refer to \cite{Zhang-AConvergentPro,Li-ARobust} for algorithmic details. We also considered implementing asymmetric shrinkage operator and trace normalization to keep the p.s.d. and trace property of the density matrix \cite{Zhang-EffRecDenMat,Zhang-AConvergentPro,Li-ARobust}.

\section{VI. CONCLUSION}
In this paper, we provided a tailored efficient framework for reconstructing mixed quantum states up to $12$ qubits from an incomplete set of observables. We applied the quantum density matrix constraints and proposed a Quantum-ADMM algorithm to accelerate the convergence. Our algorithm used $8,35$, and $226$ seconds respectively to reconstruct superposition states of $10,11,12$ qubits using $1 \%$ of measurements, which is the fastest realization to date. Experimental data of mixed states obtained in an ion trap experiment verified its effectiveness.

\section{ACKNOWLEDGMENTS}

We thank Z.K. Li for valuable discussions and A. Liutkus for sharing their codes.
This work was supported by the National Natural Science Foundation of China under Grant No. 61573330.

\section{APPENDIX}

\subsection{Rank Restricted Isometry Property}
\begin{deft}[$\mathbf{ Rank \ RIP}$]\cite{RECHT-Guaranteed,Liu-Universal-Pauli}
The $\mathcal{A}$ satisfies the rank restricted isometry property (RIP) if for all $d \times d$ $\mathbf{X}$, we have
\begin{equation}
(1-\delta)||\mathbf{X}||_F \leq ||\mathcal{A} (\mathbf{X})||_2 \leq (1+\delta)||\mathbf{X}||_F
\end{equation}
where some constant $0 < \delta <1$.
\end{deft}

\subsection{Alternating Direction Multiplier Method (ADMM)}
An optimization method to solve problems with two objective functions: $\min{ f(x) + g(z)}, \text{s.t.} \mathbf{A}x + \mathbf{B}z  = c$ where $x,z \in \mathbf{R}^N$ are variables, $\mathbf{A} \in \mathbf{R}^{P\times N}, \mathbf{B} \in \mathbf{R} ^{P\times M}, c \in \mathbf{R}^p$, $f$ and $g$ are two convex functions. Generally, ADMM iterates can be written as follows
\begin{equation}
\left\{ \begin{array}{c} x^{k+1} = \arg \min_x \{ f(x) + \lambda/2||\mathbf{A}x + \mathbf{B} z^{k} - c +b^k/\lambda||^2_2\}  \\ z^{k+1} = \arg \min_z \{ g(z) + \lambda/2 || \mathbf{A}x^{k+1} + \mathbf{B}\-c + b^k/ \lambda||_2^2\} \\ b^{k+1} = b ^k + \kappa \lambda(\mathbf{A} x^{k+1} + \mathbf{B} z^{k+1} -c) \end{array}\right.
\end{equation}
where $b \in {\mathbf{R}^M}$ is the Lagrangian multiplier, $\lambda>0$ is the penalty parameter, $\kappa >0$ is a convergence parameter.

\bibliographystyle{apsrev4-1} 

%

\end{document}